# Speeding Up Computers

Janusz Kowalik, Piotr Arłukowicz The University of Gdansk, Poland
and Erika Parsons The University of Washington

There are two distinct approaches to speeding up large parallel computers. The older method is the General Purpose Graphics Processing Units (GPGPU). The newer is the Many Integrated Core (MIC) technology . Here we attempt to focus on the MIC technology and point out differences between the two approaches to accelerating supercomputers. This is a user perspective.

## Introduction and historic background

For over thirty years until 2004-5 the speed of processors followed the Moore's Law and was exponentially increasing. Approximately every 18 months processors speed doubled. This rapid and steady progress was very economically satisfying. It was enough to wait several months for new processor generation without the need for redeveloping software. However shortly after the electronic technology hit energy consumption and heat dissipation walls in the middle of the new century first decade we witnessed the appearance of multi-core processors. Initially with two or four cores. Later up to eight. The growth of processor speed continued but the technology behind the increased performance was parallelism not the higher clock frequency. A more dramatic change took place when NVIDIA and AMD designed and manufactured highly parallel processing devices with many cores. Initially they were used mainly for graphics applications and called Graphics Processing Units (GPU). Later they have become useful for general purpose applications (GPGPU) beyond graphics and visualization [Kirk2010, Sanders2011].

Until the end of 2012 GPU processors dominated the field of High Performance Computing without a rival technology. Several large supercomputers used GPUs as accelerators. For example, the TOP500 champion in November 2012 the Cray TITAN supercomputer at Oak Ridge National Laboratory was accelerated by NVIDIA Kepler GPUs.

The HPC scene changed again at the end of 2012 when the Intel Many Integrated Core (MIC) architecture's first product the Xeon Phi coprocessor appeared

[Jeffers2003, Rahman2013]. MIC was preceded by several exploratory projects. One of them was the research project Many Core Larrabee.

These early projects were sufficiently successful to allow Intel in 2011 building and delivering to the Texas Advanced Applications Computing Center (TAACC) the 8 PetaFlops supercomputer called Stampede. The MIC Xeon Phi processor was officially announced in 2012 at the International Supercomputing Conference in Hamburg Germany Fig. 1. The many integrated core Xeon Phi coprocessor may be regarded as a remote cousin of GPU processors because it is highly parallel and can be used to accelerate general purpose computing but Phi is based on significantly different hardware design and programming principles. Here we focus on programming and performance of this newest HPC player. Xeon Phi may greatly influence the future HPC developments on the path to the exascale computation goal expected to happen between 2018 and 2020.

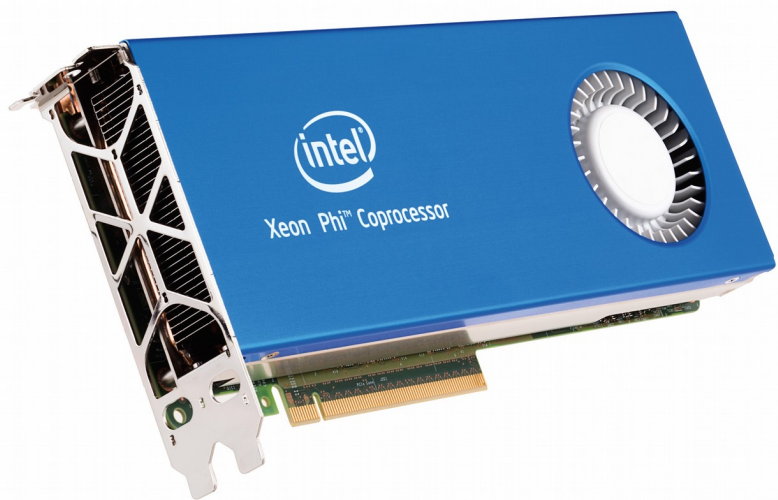

Fig. 1. Intel® Xeon Phi™ coprocessor. More information and specification: http://ark.intel.com/products/series/71842 (access date 20160215).

# Additional information.

A good source of information how to program Xeon Phi coprocessor is the book [Jeffers2003] by Jim Jeffers and James Reinders "Intel Xeon Phi Coprocessor High Performance Programming" published by Elsevier and Morgan Kaufmann in 2013. Another helpful reference is [Rahman2013].

# Briefly about Phi architecture.

The simplest definition of the Intel Xeon Phi coprocessor architecture is Symmetric Multi-Processor (SMP) on-a-chip.

A slightly more accurate description would be ccUMA SMP on-a-chip, that would indicate two key properties of this parallel computer: cache coherence of Phi coprocessor and its Uniform Memory Access property contrasting with the distributed shared memory ccNUMA architectures. Fig. 2 shows simplified architecture of coprocessor Phi.

All cores are connected by a bidirectional ring Fig. 2. The initial experience indicates that the memory access time is uniform hence the name ccUMA is justified.

It is important to stress that each core has a scalar processing unit and a vector processing unit called VPU.

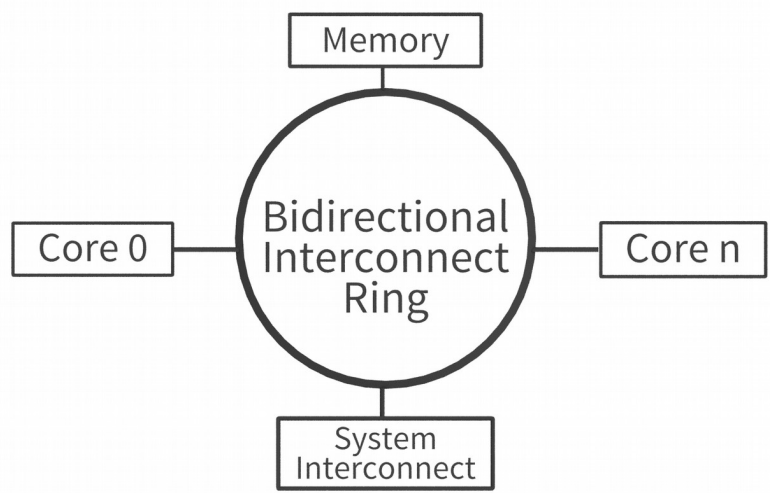

Fig. 2. Simplified coprocessor Phi architecture

The name coprocessor will be explained in the section on the Phi execution options. Since Xeon Phi coprocessor cannot function alone without a host CPU the smallest configuration of a heterogeneous computer with Xeon Phi contains one CPU and one coprocessor. More interesting is the fact that coprocessors Phi and CPUs may be clustered and serve as building blocks for constructing large scale powerful supercomputers. A good example is the June 2013 champion of the TOP500 list the Chinese supercomputer called Tianhe 2 using the Xeon Phi coprocessors for speeding up the huge machine.

## Performance

Intel has three families of Xeon Phi products. We limit our description to Phi coprocessor from the family designated as 5100. Table 1 contains specifications of Xeon Phi coprocessor. Xeon Phi has 61 cores whose clock frequency is lower than the clock frequency of the current top Xeon multi-core processors.

| Intel® Xeon Phi™ Coprocessor Specification | |
|---|---|
| Processor base frequency in GHz | 1.091 |
| Number of cores | 61 |
| Memory size and type | 8 GB DDR5 |
| Memory Speed in GT/sec | 5.5 |
| Double Precision Peak Memory Speed in Tflop/s | 1.065 |
| Single Precision Peak Memory Speed in Tflop/s | 2.1300 |
| Peak Memory Bandwidth in GB/s | 352 |

Table 1. Xeon Phi coprocessor specifications, published by Intel in [Jeffers2003]. Note: Specifications for the multiple available production models will vary.

The advertised peak performance for double precision is slightly above 1 TeraFlop/s. The single precision speed is two times faster. What is important to note is that this maximal performance can be achieved in benchmark runs and in real applications. Before discussing an example of Phi performance test we may ask a general question: when and how should we use Xeon Phi coprocessor. A simple chart in Fig. 3 provides a basis for answering this question.

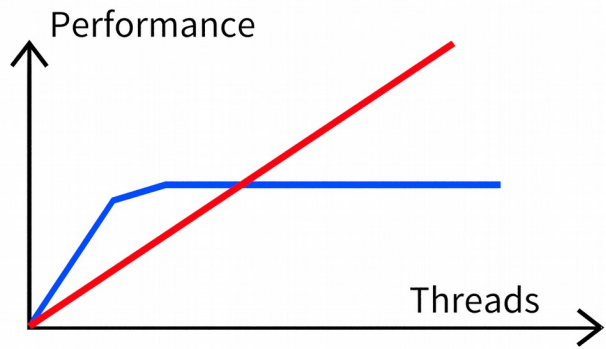

Fig. 3. Performance of Phi (red) and CPU processor (blue).

Using the Xeon Phi would benefit significantly computation if the number of used threads is sufficiently high. For small size problems with very limited parallelism it is more beneficial using faster multi-core processors. Each Phi core can run up to four threads. The minimal required number of Phi threads is application dependent but the number of 100 threads has been recommended by pioneer users of Phi coprocessor. The chart also suggests that if algorithm is a mix of highly parallel parts and sequential or lowly parallel parts both processors should share the computational effort for the best performance.

The rule of minimum 100 threads implies that using all coprocessor cores with two threads per core is a right attempt for achieving Phi's high performance. Another most essential factor for achieving high performance is vectorization. Vectorization is the Single Instruction Multiple Data (SIMD) model of parallel computing where a single operation is applied in parallel to multiple elements of vectors. The simplest example would be adding two vectors or a slightly modified addition called Saxpy. Every core of Xeon Phi has the vector processing unit (VPU) that execute vector operations in parallel for multiple elements. In the single precision 16 operations are executed in parallel and in the double precision 8.

The Single Instruction Multiple Data (SIMD) operation is a special case of the Single Program Multiple Data (SPMD) programming model. SPMD is one of the most frequently used data decomposition techniques in parallel programming. Numerical linear algebra is full of SIMD vector operations. It follows that Xeon Phi has a good chance for becoming an effective tool for large scientific and engineering applications.

## Saxpy benchmark

Saxpy operation is the sum $z=a*x+y$ where **x** and **y** are vectors and a is scalar multiplier. Jeffers and Reinders [Jeffers2003] experimented with the Saxpy computation and demonstrated the importance of scaling and vectorization. The platform they used contained one Xeon Phi coprocessor with 61 cores and two Xeon processors with 8 cores each. They demonstrated that achieving full performance of Phi requires using all cores and at least two threads per core. They also showed that vectorization is key to achieving top speed. In addition Xeon Phi coprocessor offers very efficient Fused Multiply Add (FMA) operation. In the fused operation multiplication and addition are executed in one cycle. Such pairs of operations are common in linear algebra. For instance, they are present in: saxpy, vector dot product, matrix/matrix multiplication and matrix/vector multiplication.

## Performance Conclusions.

For modern many-core processors such as Phi parallelism is the top consideration. For Phi the number of threads should be at least 100. In general, it is recommended to run at least two threads per core for getting near optimal performance.

Intel researchers stress that vectorization is the second most important factor influencing performance. It has to be taken into account when selecting appropriate numerical algorithm solving a given problem. Algorithms involving more linear algebra computations are preferred for implementing on platforms with Xeon Phi coprocessors. In addition to parallel scaling and vectorization there are other programming rules that enhance performance of platforms with Phi coprocessors. They include:

a) Cache reuse , minimizing memory access and prefetching.

b) Minimizing the ratio of communication to computation in MPI programs.

c) Avoiding frequent transfer of data between processors and coprocessors connected by the PCI express bus.

# Three ways of using Phi in platforms consisting of CPUs and Xeon Phi coprocessors.

In a heterogeneous computer with a GPU device the relationship between CPU and GPU device can be described as the master-slave relation. The GPU device is managed by CPU. In contrast in the Intel MIC architecture processors and coprocessors are peers. Both run Linux Operating System and can either cooperate or work independently. The Xeon Phi independence and its ability to do network communication justifies the name the Phi coprocessor. Phi is not an accelerator. Making processor and coprocessor peers allows several options for using platform that contains Phi coprocessor as shown in Fig. 4.

### A. Offload option

This is the most common mode of hybrid computing using Phi. The program is launched and run on the host CPU but some selected portions of the code are offloaded for execution to the coprocessor. This option should be used if the processed algorithm has highly parallel parts that well encapsulate computing and related data. Ideally the data related to the offloaded program components should be transferred from CPU to the coprocessor only once without the need for multiple transfers. In this option the coprocessor serves as an accelerating device similar to GPU or FPGA. The code is compiled for the host processor. When offload directives are encountered and the coprocessor is running and available, the required data and code are transferred from CPU to Phi.

### B. Symmetric option

In this option the code is launched and executed on both processor and coprocessor with some workload sharing made possible by communication. An example of this option is a parallel code using MPI that engages processor and coprocessor i. The power of all available cores is used but the programmer faces two challenges. Firstly, balancing the workload since the CPU cores and the coprocessor cores are different. Secondly, the ratio of communication to computation should be low enough to minimize communication penalty.

## C. Coprocessor alone, the native option.

In this option the code is launched and executed by the coprocessor alone. Due to fast ring interconnect the issue of communication between MPI processes is less significant. However in this option not all platform computing power is used. Another potential issue is the smaller coprocessor memory than the processor RAM.

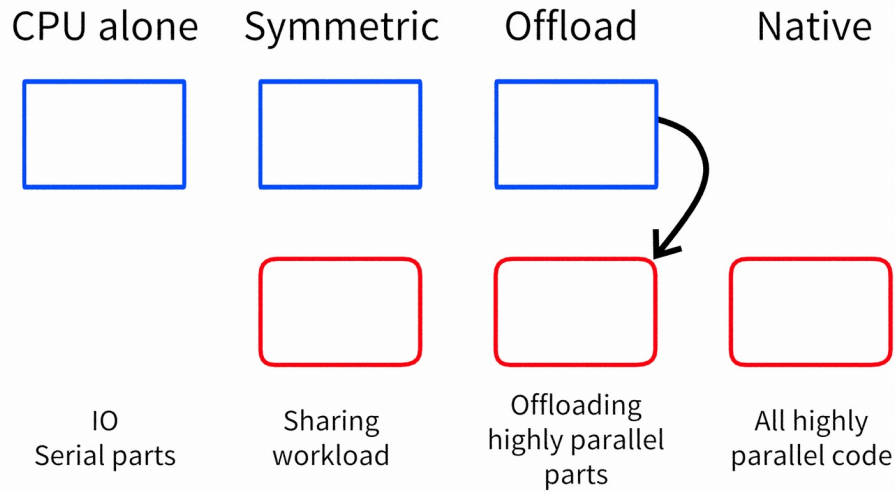

Fig. 4. Three execution options for Phi coprocessor.

The native option uses massive parallelism of Phi. However not many applications are uniformly highly parallel without sequential and lowly parallel components.

The most attractive and practical option is offload because many algorithms in scientific and technical applications are mixes of serial, lowly parallel and highly parallel parts. Offload can use CPU and coprocessor Phi for processing parts where they are best. There are two cases of offload:the non-shared memory and the virtual shared memory shown in Fig 5. The non-shared memory case applies to computation whose data are bitwise copyable such as arrays and scalars. More complex data structures such as linked lists and trees require the virtual memory.

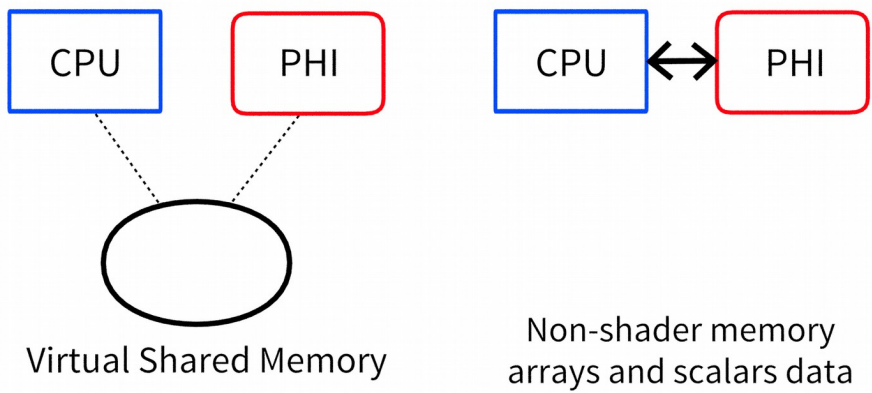

Fig. 5. Two offload cases.

## Portability.

There is significant difference between porting existing MPI and OpenMP parallel software to GPUs and the MIC technology shown in Fig. 6.

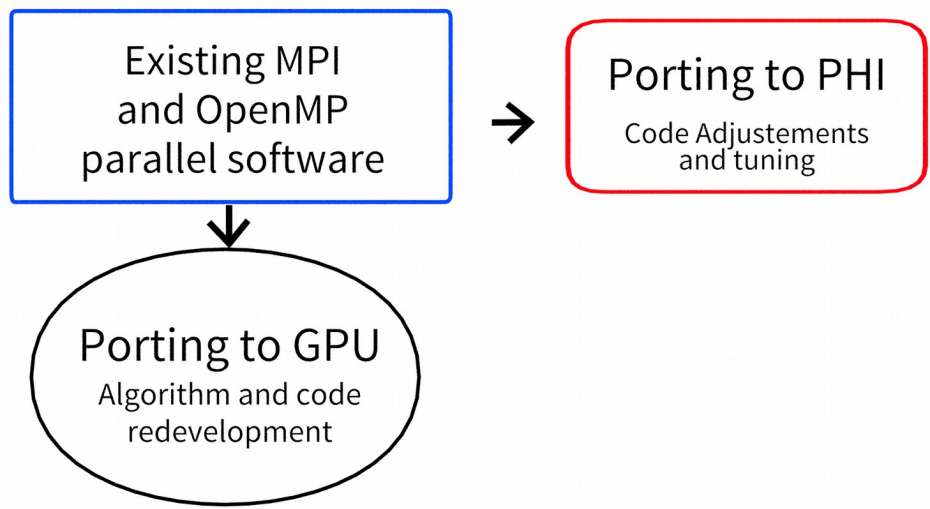

Fig. 6. Porting Difference Between Phi and GPUs.

Some new Phi users who tried porting their MPI and OpenMP programs to computers with Phi coprocessors called the porting effort trivial. No porting is really trivial in HPC where optimal performance is often required. But some cases of porting are easier and less expensive than others. For instance, porting a large MPI or OpenMP code to a GPU programmed by CUDA is harder and longer than porting it to a platform with Xeon Phi coprocessors.

The current Xeon Phi supports a broad range of languages and parallel programming tools. They include: languages like C, C++, Fortran, Coarray Fortran, OpenCL [Kowalik2012] and systems MPI [Gropp1994], OpenMP [Chapman2008], Intel Cilk, Intel TBB and the Intel Math Kernel Library (MKL).

It is conceivable that CUDA and similar GPU related programming systems will succeed as methods for programming devices for special application niches such as graphics and visualization that are naturally highly parallel and where these systems have developed solid application bases. It might be harder for the GPU systems to compete in the application areas where MPI and OpenMP already have succeeded and established acceptance as the standard tools for developing HPC software. Here economics of porting would be more likely in favor of Xeon Phi. Developing MIC and its predecessors Intel correctly avoided proposing new parallel languages. Instead it extended and modified the familiar languages allowing the use of successful existing standard programming tools such as MPI and OpenMP. MPI and OpenMP together cover two mainstream computer architectures, clusters and shared memory computers. This way Xeon Phi is riding on the two major parallel programming standard systems that have proved already their practical value for industrial and scientific applications. It should be added for fairness that GPU accelerators are massively parallel and offer impressive speedups. Once proper software is developed GPUs offer great speeds required in HPC applications.

## Xeon Phi in the TOP500 list of computers.

Since June 2013 the champion of the TOP500 world fastest supercomputers has been the Chinese huge supercomputer called Tianhe 2 (Milky Way 2). It is based on the Intel MIC technology and contains 48000 Xeon Phi boards. All coprocessors have 57 cores each. The total number of cores was 3120000.

The second machine from the top was Cray XK7 Titan with NVIDIA K20 devices.

The sixth place took the supercomputer called Stampede accelerated by the Intel Xeon Phi coprocessors.

In general, in 2014 17 out of all TOP500 computers used the Intel Xeon Phi processors. This number increased to 35 in 2015.

## Phi coprocessor solving systems of linear algebraic equations.

Linear algebra is the working horse of applied mathematics and linear equation solvers are important part of linear algebra methods.

One of the preferred methods for solving linear algebraic equations is the Conjugate Gradients Method (CGM). Linear systems of algebraic equations arise often in partial differential equations PDE) problems solved by using finite element or finite difference methods. The systems matrices are large and sparse. CGM can be used as an alternative to the Choleski decomposition for systems with symmetric positive definite matrices. A significant advantage of CGM over the decomposition is its use of the original sparse matrix in every iteration. In decomposition methods we must use special sparse matrix techniques that limit growth of non zeros and ensure numerical stability. CGM is numerically stable.

In addition to its applicability to many scientific and engineering problems CGM is mathematically elegant, numerically stable and computationally attractive because it uses only three common linear algebra operations: matrix-vector multiplication, vector dot product and Saxpy described earlier. The method applies to systems of linear equations with positive definite matrices that often arise in scientific and engineering applications. The algorithm is iterative but it behaves like a direct method because it can be shown that using exact arithmetic the method converges in n iterations for a system with n equations. The method has several versions that improve its practical performance. More information about matrix computations can be found in a classic linear algebra book [Golub1990].

The basic Conjugate Gradient Method algorithm is shown in Fig. 7.

```
              (x⁽⁰⁾ϵRⁿ given)
1.      x = x⁽⁰⁾
2.      r = b-Ax
3.      p = r
4.      α = ‖r‖²
5.      while α > tol²:
6.          λ = α / (pᵀAp)
7.          x = x + λp
8.          r = r - λAp
9.          p = r + (‖r‖²/α)·p
10.         α = ‖r‖²
11.     end
```

Fig. 7. Basic Conjugate Gradient Method.

**Notation:** Greek letters are scalars. Roman letters are column vectors except the linear system matrix **A**. The upper script **T** means transpose.

## CGM algorithm.

Lines 1, 2 and 3 create the initial solution vector x, vectors r and p. Lines 4 and 5 serve as the stopping criterion that terminates iterations if CGM converged. The remaining lines 6 to 10 define the iterative process of CGM.

Every iteration has: one matrix/vector multiplication, two vector dot products and three Saxpy operations.

The CGM algorithm solves the equations **Ax=b** where **A** is a positive definite matrix. If initially **A** is not positive definite we can multiply both sides of the equation **Ax=b** by the transpose of **A** and obtain an equivalent system of equations with the required property of the new system matrix. This conversion has a downside since the condition number of the new matrix is the square of the condition of **A**. But this difficulty may be repaired by some preconditioning technique [Golub1990].

**The platform with coprocessors Phi is ideally suited for solving large systems of linear equations by the CGM method whose numerical algorithm is fully vectorizable**.

## HPCG benchmark.

Ranking HPC computers has been an important scientific and professional activity. Traditionally the Linpack program for solving dense linear algebraic equations has been used as benchmark. A significant change occurred in 2014 when a new benchmark called HPCG (High Performance Conjugate Gradient) was added [Dongarra2016]. The new benchmark well supplements Linpack. Both benchmarks together better measure real performance of HPC computers for many scientific and engineering applications. This new addition allows testing sparse matrix storage and related computation techniques for large matrices that arize from PDEs. The authors mention that HPCG rewards efficient programming such as using collective communication in MPI and local memory access essential for large computers with many local memory domains. Reference [Dongarra2016] provides specific performance results that include machines with Xeon Phi coprocessors.

More information can be found in www.hpcg-benchmark.org

Contact with the authors:

j. kowalik@comcast. net
piotao@gmail.com
efuente@uwb. edu